\documentclass[RNAAS]{aastex62}

\usepackage{color}

\begin{document}

\title{Two nested shells around the blue supergiant ALS\,19653}

\correspondingauthor{J.M. Drudis}
\email{astrodrudis2016@gmail.com}

\author[0000-0002-2526-2758]{J.M. Drudis}
\affiliation{Curiosity2 Observatory, New Mexico Skies, Mayhill, NM 88339, USA}
\author[0000-0003-1536-8417]{V.V. Gvaramadze}
\affiliation{Sternberg Astronomical Institute, Lomonosov Moscow State University, 
Universitetsky pr., 13, Moscow 119234, Russia}
\affiliation{Space Research Institute, Russian Academy of Sciences, Profsoyuznaya 84/32, 
117997 Moscow, Russia}




\section{Abstract}
We present the results of deep narrowband imaging of two nested shells around the blue supergiant 
ALS\,19653, which confirm that the outer shell is physically associated with the star. 

\section{Introduction}  

Massive stars experience episodes of intense mass loss during their lifetime, leading to the formation of 
parsec-scale circumstellar nebulae of various forms. Searching for such nebulae serves as an effective way 
to detect rare types of massive stars, such as the luminous blue variables, blue supergiants and Wolf-Rayet 
stars (e.g. Gvaramadze et al. 2010, 2012). One of the massive stars revealed in this way is the central 
star, ALS\,19653, of a dumbbell-like infrared nebula discovered (Gvaramadze et al. 2019, hereafter Paper\,I) 
in the archival data of the {\it Wide-field Infrared Survey Explorer} ({\it WISE}; Wright et al. 2010). 
Follow-up inspection of H$\alpha$+[N\,{\sc ii}] images from the Isaac Newton Telescope (INT) Photometric 
H$\alpha$ Survey of the Northern Galactic Plane (IPHAS; Drew et al. 2005) lead to the detection of an almost 
circular shell (of angular radius of $\approx0.5$ arcmin or $\approx0.2$\,pc) surrounding the infrared nebula 
and a more extended and faint shell-like structure (of angular radius of $\approx2.7$ arcmin or 
$\approx1.2$\,pc) also centered on ALS\,19653 (Paper\,I). Subsequent optical spectroscopic observations 
showed that ALS\,19653 is a B0.5 supergiant, and allowed to derive the expansion velocities of the inner and 
outer shells of $\approx20-30$ and $100 \, {\rm km} \, {\rm s}^{-1}$, respectively (Paper\,I). It was also 
found that the spectrum of the inner shell shows only emission lines of H$\alpha$ and [N\,{\sc ii}] 
$\lambda\lambda$6548, 6584, which is consistent with the low expansion velocity of this shell. On the other 
hand, the higher expansion velocity derived for the outer shell implies that this shell could be a source of 
[O\,{\sc iii}] emission. The main purpose of the study presented in this note is the deep narrowband imaging 
of the environs of ALS\,9653, aimed to get a better idea of the morphology of the outer shell and to confirm 
its connection to the star.

\section{Narrowband imaging} 

To fulfill our purpose, a deep color image was planned to be taken with the 61 cm telescope (f/6.5; 
camera FLI PL16803 16MPx, 9$\mu$) at the Curiosity2 Observatory using narrowband (FWHM 15\,\AA) H$\alpha$, 
[N\,{\sc ii}] and [O\,{\sc iii}] filters centered, respectively, at 6563, 6584 and 5007 \,\AA. The exposure 
times for individual images taken with these filters were set at 40 min each. After taking 37 H$\alpha$ and 
26 [O\,{\sc iii}­] images (total exposure times of $\approx25$ and 17 hours, respectively), it became 
clear that, on one side, our H$\alpha$ image (Figure\,1(b)) did not lack any detail that was already 
present in the IPHAS image. This was a strong indication that there was no specific [N\,{\sc ii}] 
emission in this nebula, and we then dropped the capture of any frames with the [N\,{\sc ii}] filter. 
The second conclusion was that the [O\,{\sc iii}­] images did not show any traces of neither the inner nebula 
(which is consistent with the lack of the [O\,{\sc iii}] $\lambda$5007 line in its spectrum; see Paper\,I) 
nor the outer shell. Correspondingly, for the final color image (Figure\,1(a)) we used only the H$\alpha$ 
(red) image and a set of short exposure frames through the broadband $R$, $G$ and $B$ filters in order to 
depict the stars in the image with their true natural colors. The processing of this image was very simple 
in order to preserve the original structures. The only treatment was a dual deconvolution applied on 
different parts of the image. A maximum entropy deconvolution was applied to the outer shell, in order to 
enhance its presence, and a positive constraint deconvolution was applied to the central nebula in order 
to obtain greater detail of its structure. Figure\,1(b) shows the outer shell without applying the maximum 
entropy deconvolution, in order to show its real brightness, compared to the central nebula. 

Figure\,1(a) shows that the outer shell represents an almost circular nebula whose emission fills nearly  
the entire space between its outer edge and the central nebula. One can also see that the maximum brightness 
of the shell falls on its north-east edge and that ALS\,19653 is somewhat offset towards this edge. This 
could be caused either by a north-east motion of the star relative to the ambient medium or by a density 
gradient in the same direction.

\section{Space velocity of ALS\,19653}

To check whether the space motion of ALS\,19653 is the cause of the increased brightness of the north-east
edge of the outer shell, we calculated the peculiar transverse velocity, $v_{\rm tr}$, of this star using 
its {\it Gaia} DR2 proper motion ($\mu _\alpha \cos \delta=-0.188\pm0.113 \, {\rm mas} \, {\rm yr}^{-1}$, 
$\mu _\delta=-2.203\pm0.112 \, {\rm mas} \, {\rm yr}^{-1}$) and parallax ($0.639\pm0.063$ mas) measurements \citep{Gaia2018}. To do this, we adopted the solar Galactocentric distance of $R_0 =8.0$ kpc and the 
circular Galactic rotation velocity of $\Theta _0 =240 \, {\rm km} \, {\rm s}^{-1}$ \citep{Reid2009}, and 
the solar peculiar motion of $(U_{\odot},V_{\odot},W_{\odot})=(11.1,12.2,7.3) \, {\rm km} \, {\rm s}^{-1}$ 
\citep{Schonrich2010}. We derived the peculiar velocity components along the Galactic longitude and latitude 
of $v_{\rm l}=8.5\pm0.8 \, {\rm km} \, {\rm s}^{-1}$ and $v_{\rm b}=1.4\pm0.8 \, {\rm km} \, {\rm s}^{-1}$, 
respectively, which correspond to $v_{\rm tr}=8.6\pm0.8 \, {\rm km} \, {\rm s}^{-1}$. Using the systemic 
radial velocity of ALS\,19653 of $-9.0\pm0.1 \, {\rm km} \, {\rm s}^{-1}$ (Paper\,I), we calculated also the 
peculiar radial velocity $v_r=-17.6\pm0.1 \, {\rm km} \, {\rm s}^{-1}$ and the total space velocity of this 
star of $v_{\rm tot}=19.5\pm0.36 \, {\rm km} \, {\rm s}^{-1}$. The quoted uncertainties in the velocity 
estimates take into account only the uncertainties in the proper motion and systemic velocity measurements. 
Although the derived value of the space velocity did not allow us to treat ALS\,19653 as a classical runaway 
star, its orientation in the sky shows that this star is moving in the ``correct" (north-east) direction, 
meaning that the enhanced brightness of the north-east edge of the outer shell could indeed be caused by this 
motion. 

\section{Discussion}

Detection of two nested shells around ALS\,19653 indicates that this star at least twice went through 
episodes of enhanced mass lose in the recent ($\sim10^4$\,yr) past. Similar two-component nebulae 
were also found around both massive and low-mass stars, e.g., the O6f?p star HD\,148937 
\citep{Leitherer1987} and the M1\,II star HD\,65750 \citep{Drudis2018}. The common characteristic of these 
objects is that their inner nebulae possess axial symmetry. Although the mechanisms that lead to the formation 
of such nebulae are not fully understood, it is believed that they are somehow related to the present or past 
binarity of their central stars. In particular, the fast rotation and strong magnetic field of HD\,148937 and 
the nitrogen-rich composition of its inner nebula make this star a strong merger candidate \citep{Langer2012}. 
Similarly, it was suggested that ALS\,19653 is a merger product as well (Paper\,I). The possible binarity 
of these two stars \citep{Wade2019,Gvaramadze2019} does not contradict these suggestions because both stars 
could originally be members of triple or higher multiple systems \citep[cf.][]{Langer2012}. 

Finally, we note that a careful inspection of the {\it WISE} 22\,$\mu$m image revealed that the central 
nebula is surrounded by a diffuse halo of the same angular extent as the outer optical shell (Figures\,1(c)
and 1(d)), which provides further evidence of the connection of this shell with ALS\,19653. 
 
V.V.G. acknowledges support from the Russian Foundation for Basic Research grant 19-02-00779.

 \begin{figure}
 \begin{center}
 \includegraphics[scale=0.17,angle=0]{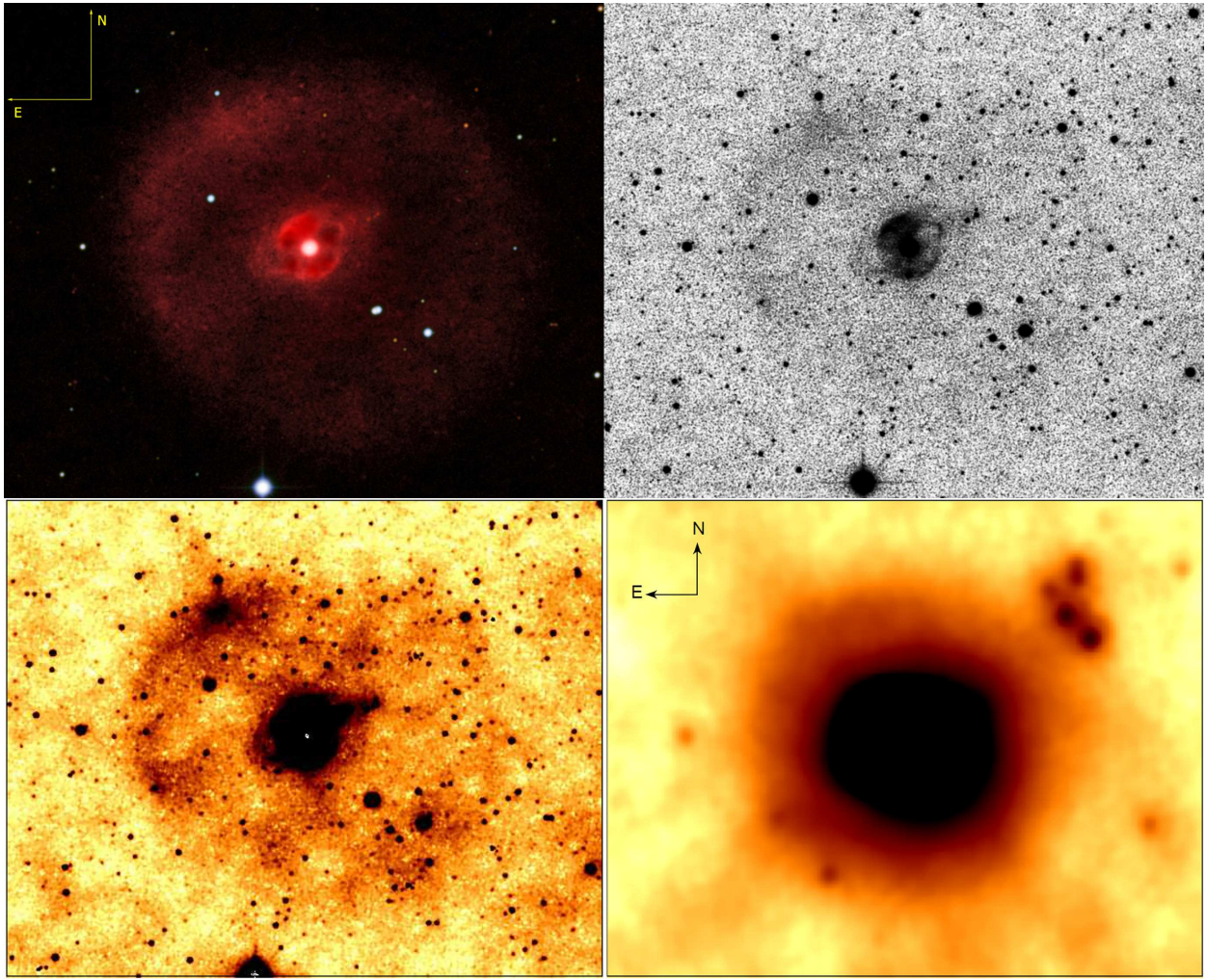}
 \caption{Figure\,1(a) (upper left): final color image. Figure\,1(b) (upper right): inverted H$\alpha$ 
 image. Figures\,1(c) and 1(d) (bottom panels): H$\alpha$ (left) and {\it WISE} 22\,$\mu$m (right) images. 
 The orientation and the scale of all images are the same.
 \label{fig:neb}}
 \end{center}
 \end{figure}

\end{document}